\documentclass[aps,prl,reprint,superscriptaddress,amsmath,footinbib]{revtex4-1}
\usepackage{graphicx}
\usepackage{bbold}
\usepackage{hyperref}
\usepackage{import}
\usepackage{color}
\usepackage{soul}
\usepackage{mathrsfs}
\usepackage{amsbsy}
\usepackage{xcolor}

\newcommand{\bra}[1]{\ensuremath{\langle #1|\,}}

\newcommand{\ket}[1]{\ensuremath{\,|#1\rangle}}
\newcommand{\Ca}{\ifmmode ^{40}\text{Ca}^{+} \else $^{40}$Ca$^{+}$~\fi}
\newcommand{\mus}{\ifmmode \mu\mathrm{s} \xspace \else $\mu$s \xspace\fi}

\begin{document}


\title{Enhanced Quantum Interface with Collective Ion-Cavity Coupling} 

\author{B. Casabone}
\affiliation{Institut f{\"u}r Experimentalphysik, Universit{\"a}t Innsbruck, Technikerstra{\ss}e 25, 6020 Innsbruck, Austria}
\author{K. Friebe}
\affiliation{Institut f{\"u}r Experimentalphysik, Universit{\"a}t Innsbruck, Technikerstra{\ss}e 25, 6020 Innsbruck, Austria}
\author{B. Brandst{\"a}tter} 
\affiliation{Institut f{\"u}r Experimentalphysik, Universit{\"a}t Innsbruck, Technikerstra{\ss}e 25, 6020 Innsbruck, Austria}
\author{K. Sch{\"u}ppert}
\affiliation{Institut f{\"u}r Experimentalphysik, Universit{\"a}t Innsbruck, Technikerstra{\ss}e 25, 6020 Innsbruck, Austria}
\author{R. Blatt}
\affiliation{Institut f{\"u}r Experimentalphysik, Universit{\"a}t Innsbruck, Technikerstra{\ss}e 25, 6020 Innsbruck, Austria}
\affiliation{Institut f{\"u}r Quantenoptik und Quanteninformation, {\"O}sterreichische Akademie der Wissenschaften, Technikerstra{\ss}e 21a, 6020 Innsbruck, Austria}
\author{T. E. Northup}
\affiliation{Institut f{\"u}r Experimentalphysik, Universit{\"a}t Innsbruck, Technikerstra{\ss}e 25, 6020 Innsbruck, Austria}
\date{\today}

\begin{abstract}
We prepare a maximally entangled state of two ions and couple both ions to the mode of an optical cavity. 
The phase of the entangled state determines the collective interaction of the ions with the cavity mode, that is, whether the emission of a single photon into the cavity is suppressed or enhanced. 
By adjusting this phase, we tune the ion--cavity system from sub- to superradiance. 
We then encode a single qubit in the two-ion superradiant state and show that this encoding enhances the transfer of quantum information onto a photon.
\end{abstract}  
\maketitle 

Sub- and superradiance are fundamental effects in quantum optics arising in systems that are symmetric under the interchange of any pair of particles~\cite{Dicke54,Gross82,Garraway11}.
Superradiance has been widely studied in many-atom systems, in which effects such as a phase transition~\cite{Baumann10,Baden14} and narrow-linewidth lasing~\cite{Bohnet12} have recently been observed.
For few-atom systems, each atom's state and position can be precisely controlled, and thus collective emission effects such as Rydberg blockade \cite{Lukin01} and the Lamb shift \cite{Meir13} can be tailored.  
In a pioneering experiment using two trapped ions, variation of the ions' separation allowed both sub- and superradiance to be observed, with the excited-state lifetime extended or reduced by up to 1.5\% ~\cite{DeVoe1996}.
The contrast was limited because spontaneous emission from the ions was not indistinguishable, as the ions' separation was on the order of the wavelength of the emitted radiation. 
{This limitation can be overcome by observing preferential emission into a single mode, such as the mode defined by incident radiation \cite{Dicke54} or by an optical cavity.
In a cavity setting, indistinguishability is guaranteed} when the emitters are equally coupled to the mode, even if they are spatially separated. 
Subradiance corresponds to a suppressed interaction of the joint state of the emitters with the cavity mode, while for the superradiant state, the interaction is enhanced.  

In the context of quantum networks~\cite{Kimble08a,Duan10}, superradiance can improve a 
{ 
quantum interface when one logical qubit is encoded across $N$ physical qubits.
In the DLCZ protocol for heralded remote entanglement, efficient retrieval of stored photons is based on superradiance~\cite{Duan01,Oliveira14}.
Superradiance can also improve the performance of a deterministic, cavity-based interface, which enables the direct transmission of quantum information between network nodes~\cite{Cirac97}. 
If a qubit is encoded in the state $\frac{1}{\sqrt{N}}\sum_i^N|\downarrow_1...\uparrow_i...\downarrow_N\rangle$, the coupling rate to the cavity is enhanced from the single-qubit rate $g$ to the effective rate $g \sqrt{N}$, relaxing the technical requirements for strong coupling between light and matter~\cite{Lamata11}.
This state corresponds to the first step in the superradiant cascade described by Dicke~\cite{Dicke54}.
In contrast, subradiant states are antisymmetrized, resulting in suppressed emission.
From a quantum-information perspective, subradiant states are interesting because they span a decoherence-free subspace \cite{Plenio99,Beige00,Lidar03}.}
A subradiant state of two superconducting qubits coupled to a cavity has recently been prepared \cite{Filipp11}.

\begin{figure}[h]
\includegraphics[width=0.475\textwidth]{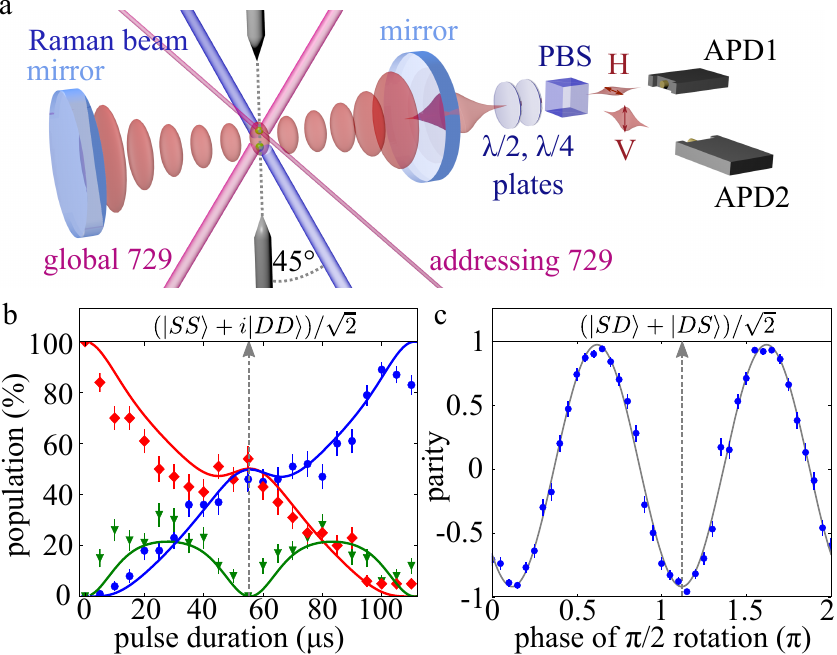}
\caption{(a) Two \Ca ions in a linear Paul trap couple with equal strength to the mode of a high-finesse optical cavity. 
A magnetic field orthogonal to the cavity axis 
defines the quantization axis. 
Quantum information stored in the ions is manipulated using two 729~nm beams: the global beam couples to both ions, while the addressing beam is focused onto one ion. 
A 393~nm laser beam 
drives a cavity-mediated Raman transition, generating a single photon in the cavity. 
At the cavity output, two wave plates ($\lambda/2$, $\lambda/4$) select the basis in which photon polarization is analyzed.
Two avalanche photodiodes (APD1 and APD2) detect the horizontally (H) or vertically (V) polarized photons at the output of a polarizing beamsplitter (PBS).
(b) Populations of the states $\ket{S}\ket{S}$ (red diamonds), $\ket{D}\ket{D}$ (blue circles), and $ \ket{S}\ket{D}$ or $\ket{D}\ket{S}$ (green triangles) as a function of the M{\o}lmer--S{\o}rensen gate duration. 
After 55~$\mu$s (dashed vertical line) a maximally entangled state is generated.
 Solid lines indicate the ideal time evolution of the gate {operation~\cite{Benhelm08}.
(c) Oscillations in the parity of the ion populations as a function of the phase of a $\pi$/2 pulse on the $\ket S \leftrightarrow \ket D$ transition, following entanglement.
The dashed vertical line at phase $1.2\,\pi$ 
corresponds to $\ket{\Psi^+}$.
}
Error bars represent projection noise.}	
\label{fig1}
\end{figure}

Here, we generate collective states of two ions coupled to an optical cavity and use a state 
{ 
that maximizes the coupling rate}
 to improve ion--photon quantum information transfer. 
Our system is described by the Tavis--Cummings Hamiltonian~\cite{Tavis68}, the interaction term of which is
\begin{equation}
\label{TCH}
H_\text{int} = \hbar g \left( \sigma^{(1)}_{-} +  e^{i\zeta} \sigma^{(2)}_{-}\right)a^{\dagger} + \text{h.~c.},
\end{equation}
where $\sigma^{(j)}_{-}$ is the lowering operator for the $j$th ion, $\zeta$ represents a relative phase \cite{SM}\nocite{Nielsen2000, Efron93, Tan99}, and $a^\dagger$ is the creation operator of a photon in the cavity mode.
We prepare a maximally entangled two-ion state and tune its emission properties between sub- and superradiance, 
that is, between a dark state $\ket{\Psi_\text{sub}}$ and a state $\ket{\Psi_\text{super}}$ that couples with enhanced strength $g \sqrt{2}$ to the cavity.
Furthermore, we transfer quantum information from a state with enhanced emission probability 
onto a single photon and show that the process fidelity and efficiency are higher than for a single-ion qubit.

In these experiments, two $\Ca$ { separated by  $5.6~\mu$m} are confined along the axis of a linear Paul trap and coupled to an optical cavity in an intermediate coupling regime~\cite{SM}.
We position the ions so that  $g_1 \approx g_2 $, where $g_j$ represents the coupling strength of the $j$th ion to the cavity \cite{Casabone13}. 
In a cavity-mediated Raman process, each ion prepared in a state from the $4^2S_{1/2}$ manifold produces a single cavity photon \cite{Barros09}.
The process is driven both by a laser at $393$~nm detuned from the $4^2S_{1/2}-4^2P_{3/2}$ transition and by the cavity, whose detuning from the 854~nm $4^2P_{3/2}-3^2D_{5/2}$ transition satisfies a Raman resonance condition \cite{Stute12a}.
Together, laser and cavity provide the interaction term of Eq. (\ref{TCH}), in which the relative phase $\zeta$ between the ions' coupling arises from the angle between the Raman beam and the ion-trap axis \cite{SM}. 
Photons leave the cavity preferentially through one mirror and are detected on 
photodiodes (Fig.~\ref{fig1}a).

Entanglement between the ions is generated using a `global' 729~nm laser beam (Fig.~\ref{fig1}a) that couples with equal strength to both ions on the $4^2S_{1/2}-3^2 D_{5/2}$ quadrupole transition.
The target state 
\begin{equation*}
\ket{\Psi^+} \equiv \left(\ket{S}\ket{D}+\ket{D}\ket{S}\right)/\sqrt{2}
\end{equation*}
is prepared via a M{\o}lmer--S{\o}rensen gate operation followed by a $\pi / 2 $ rotation, where $\ket S \equiv \ket{4^2S_{1/2}, m_j=-1/2}$ and $\ket D \equiv \ket{3^2D_{5/2}, m_j=-1/2}$. 
{ 
In the M{\o}lmer--S{\o}rensen gate, a bichromatic field that drives blue and red motional sidebands generates a spin-dependent force, coupling the ion's motion and internal state~\cite{Sorensen99}.}
Fig.~\ref{fig1}b shows the evolution of the two-ion state populations during application of the gate. 
{ 
A maximally entangled state $\ket{\Phi}= \big (\ket S \ket S +i\ket D \ket D\big )/\sqrt{2}$ is generated for a gate duration of 55~$\mu$s. 
Subsequently, a $\pi / 2 $ rotation 
maps $\ket{\Phi}$ to $\ket{\Psi^+}$.
A lower bound of 95(2)\% on the state fidelity with respect to $\ket{\Phi}$ is determined by varying the phase of the  $\pi / 2 $ rotation and measuring the parity of the ions' populations, which oscillates as a function of phase (Fig.~\ref{fig1}c) \cite{Sackett00}.
}

\begin{figure}
	\centering 
		\includegraphics[width=0.475\textwidth]{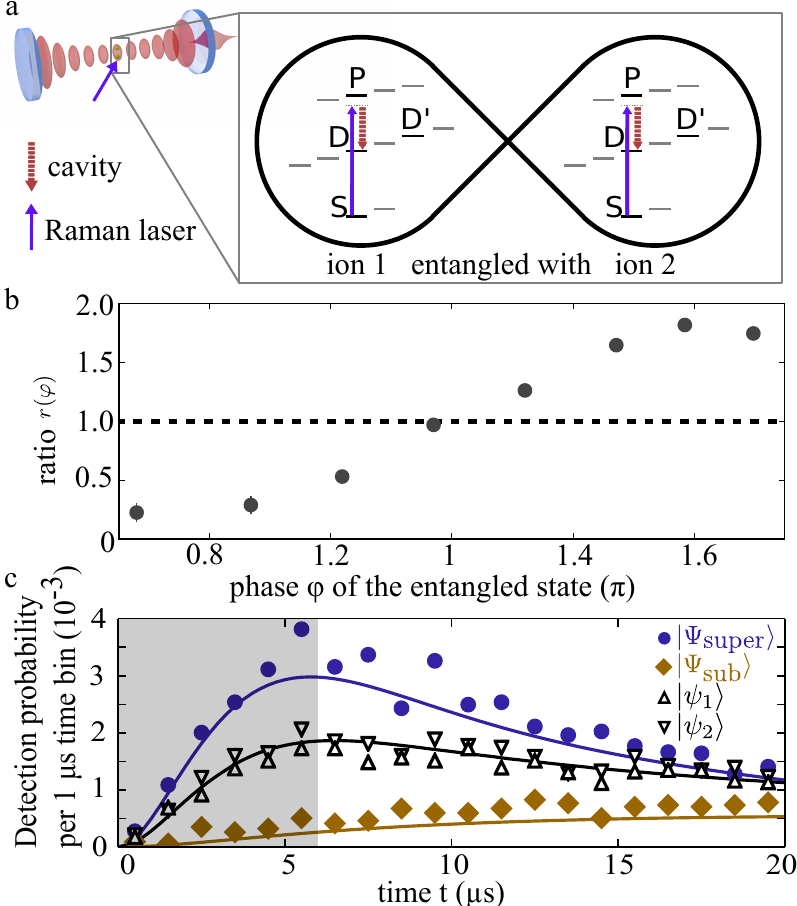}
	\caption{
(a) 
The two ions are prepared in either a separable state $\ket{\psi_1}$ or $\ket{\psi_2}$ or an entangled state $\ket{\Psi (\varphi)}$ for various values of $\varphi$.
The global beam then drives a Raman transition between $\ket S$ and $\ket D$, generating a single cavity photon for each ion in $\ket S$. Since $\ket{D'}$ is decoupled from the cavity interaction, both $\ket{\psi_1}$ and $\ket{\psi_2}$ represent a single ion interacting with the cavity. 
(b) Ratio $r(\varphi)$ of the probability to detect a photon for $\ket{\Psi (\varphi)}$ to that of $\ket{\psi_1}$ as a function of the phase $\varphi$ for the first 6~$\mu$s of the Raman process.
The reference single-ion case is shown as a dashed horizontal line.
(c) Temporal shape of the photon at the cavity output as a function of detection time $t$, for the entangled states $\ket{\Psi_{\text{super}}}$ (circles) and $\ket{\Psi_{\text{sub}}}$ (diamonds) and the single-ion cases $\ket{\psi_1}$ and $\ket{\psi_2}$ (up and down triangles, respectively). 
The temporal photon shapes are calculated by normalizing the detected photon counts  { per 1 $\mu$s time bin} by the number of photon generation attempts. 
Data are shown until 20~$\mu$s, the time scale for which enhancement and suppression are most prominent. 
Lines are simulations.
The shaded area represents the time window used in Fig.~\ref{fig2}b.
Error bars represent Poissonian statistics and are mostly smaller than the plot symbols. }
	\label{fig2}
\end{figure}

A second, `addressing' 729~nm beam with a waist smaller than the ion--ion separation couples to just one ion.
 When detuned, this beam induces AC-Stark shifts in the addressed ion, which contribute a phase $\varphi$ to the entangled state \cite{schindler13}:
 \begin{align}
 \ket{\Psi (\varphi)} \equiv \left(\ket{S}\ket{D}  + e^{i\varphi} \ket{D}\ket{S}\right)/\sqrt{2}.
 \label{eq_Psi}
 \end{align}
By adjusting the length of the Stark-shift pulse, we shift this phase, which determines the effective coupling $g_\text{eff}$ of $\ket{\Psi (\varphi)}$ to the cavity mode under the action of $H_\text{int}$.
Specifically, the superradiant and subradiant states are given by 
\begin{align}
\label{eq_super}
&\ket{\Psi_{\text{super}}} \equiv \ket{\Psi (\varphi= -\zeta)} \\
&\ket{\Psi_{\text{sub}}} \equiv \ket{\Psi (\varphi= -\zeta+\pi)}. \notag
\end{align}
Note that if $\zeta$ were zero, $\ket{\Psi_{\text{super}}}$ and $\ket{\Psi_{\text{sub}}}$ would be the Bell states $\ket{\Psi^+}$ and $\ket{\Psi^-}$, respectively.

The Raman process between $\ket S$ and $\ket D $ generates a single cavity photon from $\ket{\Psi (\varphi)}$, as only one ion is in $\ket{S}$. 
This photon has a temporal shape initially determined by 
$g_\text{eff}$ between the two-ion state and the cavity mode.
For later times, the shape is determined by the rates of both cavity decay and off-resonant scattering. 
Varying $g_\text{eff}$ by changing the phase $\varphi$ of $\ket{\Psi (\varphi)}$ thus modifies the temporal shape, that is, the probability to generate the photon early in the Raman process. 
Ideally, in the absence of scattering, the coupling of $\ket{\Psi_{\text{sub}}}$ to the cavity vanishes ($g_\text{eff}=0$) so that no photon is generated. 
For $\ket{\Psi_{\text{super}}}$, in contrast, the coupling is maximized such that $g_\text{eff}=g\sqrt{2} $. 
{ 
Thus, the probability to generate and detect a photon from $\ket{\Psi_{\text{super}}}$ early in the process is expected to be twice that of one ion.
For time scales much shorter than $1/g$, a photon generated in the cavity has not yet been reabsorbed, and therefore, cavity back-action does not play a role.
}

We now determine this probability 
for a range of phases~$\varphi$.
The experimental sequence starts with 1~ms of Doppler cooling.
The ions are then optically pumped to $\ket S$,  followed by 1.3~ms of sideband cooling on the axial center-of-mass mode~\cite{Wineland98}. 
Next, global and addressing 729~nm pulses generate the state $\ket{\Psi (\varphi)}$. 
In the last step, the cavity-mediated Raman transition is driven for 55~$\mu$s and photons are detected (Fig.~\ref{fig2}a).

In order to determine whether we achieve enhancement and suppression of the cavity coupling with respect to the single-ion rate $g$, we carry out a reference measurement.
For this single-ion case, one of the two ions is hidden in a state $\ket{D'} \equiv \ket{3^2D_{5/2}, m_j=3/2}$ that is decoupled from the Raman process.
Thus, the initial state is $\ket{\psi_{1}} \equiv \ket{S}\ket{D'}$ or $\ket{\psi_{2}} \equiv \ket{D'}\ket{S}$.

For the states $\ket{\Psi(\varphi)}$, we calculate $\eta(\varphi)$, the probability to detect a photon in the first 6~$\mu$s of the Raman process, an interval in which the effective coupling rate determines the initial slope.
For the single-ion cases, we calculate $\eta_{\psi}$, the average value of the photon detection probability for $\ket{\psi_1}$ and $\ket{\psi_2}$ in the same time window.
Fig.~\ref{fig2}b shows the ratio $r(\varphi)=\eta(\varphi)/\eta_{\psi}$ as the phase $\varphi$ is varied.
For $\varphi = 0.68 \, \pi$, the experimentally determined minimum, the ratio is 0.22(9): photon generation is strongly suppressed. 
We therefore identify $\ket{\Psi({\varphi = 0.68\,\pi})}$ with $\ket{\Psi_{\text{sub}}}$.
As $\varphi$ is increased, the ratio approaches one, then enters the superradiant regime.
A maximum value of $r(\varphi)$ is found for $\varphi = 1.58\, \pi$.  
For the corresponding state, identified with $\ket{\Psi_{\text{super}}}$, the probability to detect a photon is 1.84(4), 
{ close to its maximum value of two},  thus demonstrating strong enhancement in photon generation.

For these states $\ket{\Psi_{\text{sub}}}$ and $\ket{\Psi_{\text{super}}}$, we now analyze the temporal photon shapes at the detector (Fig.~\ref{fig2}c). 
The temporal shapes corresponding to $\ket{\psi_1}$ and $\ket{\psi_2}$ are considered as a reference; from their overlap, we find the coupling strengths of the two ions, $g_1$ and $g_2$, to be within 10\% of one another.
Photons generated from $\ket{\Psi_{\text{super}}}$ exhibit a steeper initial slope than the single-ion case, while $\ket{\Psi_{\text{sub}}}$ has a flatter slope.
The photon shapes are consistent with enhanced and suppressed coupling to the cavity and are in good agreement with simulations. 
The simulations are based on numerical integration of the master equation and include imperfect preparation of the initial state, which together 
with off-resonant scattering accounts for the small but nonzero probability to generate photons from $\ket{\Psi_{\text{sub}}}$. 
For $\ket{\Psi_{\text{super}}}$, these effects reduce the photon generation probability by about 10\% for the first 6 $\mu$s of the process~\cite{SM}. 

\begin{figure}
	\centering
		\includegraphics[width=0.475\textwidth]{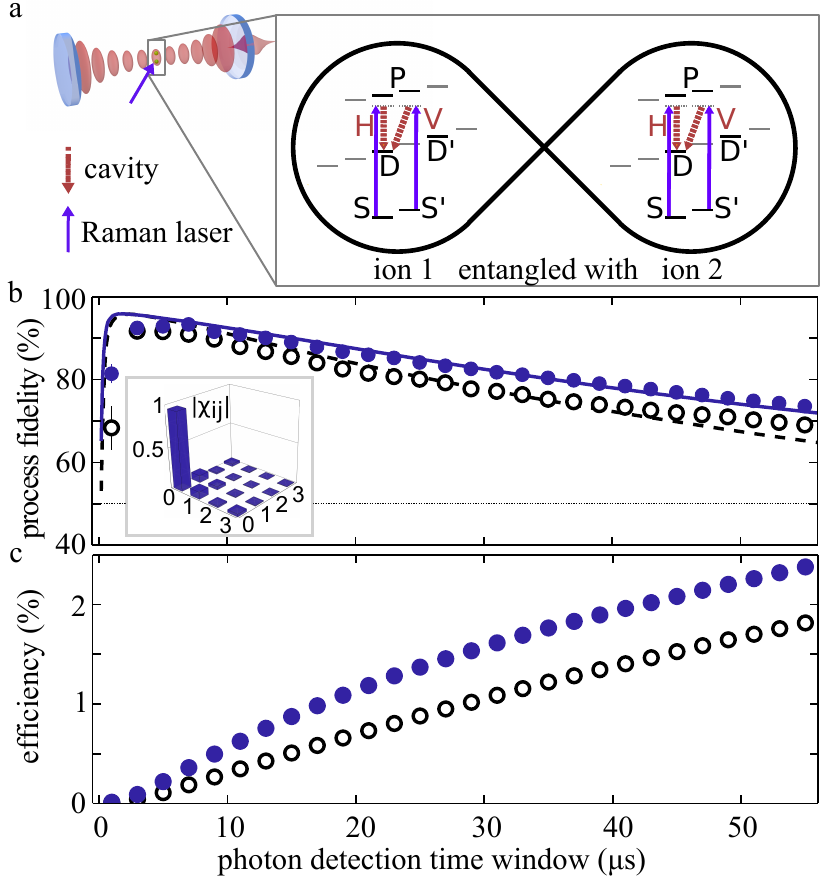}
		\caption{(a) 
A bichromatic Raman transition maps a superposition of $\ket S$ and $\ket{S'}$ onto a superposition of single-photon polarization states $\ket H$ and $\ket V$.
The superposition is encoded either in two entangled ions or in a single ion, with the other ion decoupled in $\ket{D'}$. 
	(b) Process fidelity for $\ket{\Psi_\text{super}}$ (filled blue circles) and $\ket{\psi_{1}}$ (open black circles) as a function of the photon detection time window.
Lines are simulations (continuous line: two entangled ions; dashed line: single-ion case).
Inset: absolute value of the process matrix $\chi_{ij}$ for $\ket{\Psi_\text{super}}$ reconstructed from photons detected between 2 and 4~$\mu$s, yielding the maximum process fidelity $|\chi_{00}| = 96.0(3)\%$. 
Error bars are derived from non-parametric bootstrapping.
(c) Cumulative process efficiency for $\ket{\Psi_\text{super}}$ (filled blue circles) and $\ket{\psi_{1}}$ (open black circles) as a function of the photon detection time window.
Error bars represent Poissonian statistics and are smaller than the plot symbols. }
	\label{fig3}
\end{figure}

We now describe the implementation of a quantum interface that exploits the enhanced coupling of the superradiant state to the cavity~\cite{Lamata11}.
The state $\ket{\Psi({\varphi})}$
as defined in Eq. \ref{eq_Psi} contains two contributions: 
one from the ground state $\ket{S}$ and the other from $\ket{D}$. 
We extend this definition so that the ground-state component can be stored in either of two states, that is, in $\ket{S}$ or in $\ket{S'} \equiv \ket{4^2S_{1/2}, m_j=+1/2}$.
A logical qubit is encoded in these two states, and this qubit is mapped onto the polarization state of a single cavity photon. 
To perform the mapping process, we use a phase-stable bichromatic Raman transition 
that coherently transfers $\ket{S}$ to $\ket{D}$, producing a horizontally polarized photon $\ket{H}$, and $\ket{S'}$ to $\ket{D}$, producing a vertically polarized photon $\ket{V}$~\cite{Stute13} (Fig.~\ref{fig3}a).
Defining a superposition state 
\[
	\ket{\alpha,\beta} \equiv \cos \alpha \ket S + e^{i\beta} \sin \alpha \ket {S'},
\]
the mapping process can be represented by
\begin{align}
	 &\left(\ket{\alpha,\beta}\ket{D}  + e^{i\varphi} \ket{D} \ket{\alpha,\beta}\right)  \ket{0} /\sqrt{2}  \nonumber \\
	 &\quad\mapsto \ket{D}\ket{D}   \left(\cos \alpha \ket H + e^{i\beta} \sin \alpha \ket V\right) ,
	 \label{mapping}  
\end{align}
where $\ket 0$ stands for the cavity vacuum and the phase is set to $\varphi = 1.58\pi$, corresponding to $\ket{\Psi_{\text{super}}}$.

In order to characterize the mapping, we extract the process matrix $\chi$, which describes the transformation from the input to the output density matrix: $\rho_\text{out}  = \sum_{i,j} \chi_{ij} \, \sigma_i \, \rho_\mathrm{in} \, \sigma_j$, where $\sigma_i \in \{\mathbb{1}, \sigma_x, \sigma_y, \sigma_z \}$ are the Pauli operators~\cite{Chuang97}.  
Following Doppler cooling, optical pumping, and sideband cooling as above, 
the two ions are prepared in $\ket{\Psi_\text{super}}$. 
Next, two global 729~nm pulses prepare one of the four orthogonal input states $\ket{\alpha, \beta}$, with $(\alpha, \beta) \in \{(\pi /2, 0), (0, 0), (\pi /4 ,0), (\pi/4, \pi/2)\}$.
Finally, the Raman transition is driven and the photon is detected in one of three orthogonal polarization bases ~\cite{James01}.
This set of measurements allows $\chi$ to be reconstructed via the maximum likelihood method.
As the target mapping corresponds to the identity operation, the process fidelity is given by the matrix entry $\chi_{00}$.

For comparison, 
we carry out reference measurements in which enhancement is not present, for which the ions are prepared in $\ket{\psi_1}$.
The mapping process is then given by
\begin{align}
	 \ket{\alpha,\beta}\ket{D'}  \ket{0} \mapsto \ket{D}\ket{D'}   \left(\cos \alpha \ket H + e^{i\beta} \sin \alpha \ket V\right).
	 \label{mapping}  
\end{align}

Fig.~\ref{fig3}b shows the process fidelities $\chi_{00}$ for $\ket{\Psi_\text{super}}$ and $\ket{\psi_1}$ as a function of the photon detection time window.
Not only is the fidelity of the superradiant case higher for all data points, but also the improvement over the single-ion case increases with the length of the detection window.
For a detection time window of 6~$\mu$s, the fidelity is 93.3(3)\% for $\ket{\Psi_\text{super}}$ and 90.9(5)\% for $\ket{\psi_1}$, indicating that in both cases the logical qubit is correctly mapped onto photon polarization with very high probability.
A maximum value of 96.0(3)\% is found for $\ket{\Psi_\text{super}}$ for photons detected between 2 and 4~$\mu$s.
As the detection window length is increased, $\chi_{00}$ decreases for both 
cases because the probability for off-resonant excitation to the $4^2P_{3/2}$-manifold increases with time.
If such an event happens during the Raman process, the initial state $\ket{\alpha, \beta}$ is randomly projected onto $\ket{0,0}=\ket S$ or $\ket{\pi/2,0}=\ket{S'}$, and the qubit is then mapped onto either $\ket{H}$ or $\ket{V}$, regardless of the information in the initial superposition~\cite{Stute13}.
However, while the probability for scattering is the same for both states, photons are produced earlier from $\ket{\Psi_\text{super}}$  because of the enhanced effective coupling.
Thus, the improvement in the fidelity stems from an increased probability to generate a photon before scattering occurs. 
After $55~\mu$s, we find $\chi_{00}=73.4(3)$\% for $\ket{\Psi_\text{super}}$ in comparison with 68.7(2)\% for $\ket{\psi_1}$.
Simulations that take into account detector dark counts, imperfect state initialization, { different coupling strengths of the ions to the cavity, }
and magnetic field fluctuations are in good agreement with the data.

We also investigate the cumulative process efficiency $\varepsilon(t)$, defined as the probability to detect a photon before time $t$ (Fig.~\ref{fig3}c).
For $t=6~\mu$s, the process efficiency for $\ket{\Psi_\text{super}}$ is $\varepsilon_\text{s}(t) = 0.33(1)$\%, while for $\ket{\psi_1}$, it is $\varepsilon_1 (t) = 0.17(1) $\%, corresponding to a ratio $\varepsilon_\text{s}/\varepsilon_1$ of 1.94(13).
The ratio decreases monotonically with $t$, 
and by $t=55~\mu$s, it is 1.34(5).
While the enhanced coupling modifies the temporal shape of the photons early in the process, for longer times its effect on the cumulative process efficiency is small, such that the ratio is expected to approach one.
A single photon generated in the cavity is detected with an efficiency of { 8(1)\%}, due to losses in the cavity mirrors, optical path losses and the detection efficiency of the avalanche photodiodes. 

{ 
The enhanced fidelity and efficiency of quantum state transfer in the superradiant regime can be understood in terms of a stronger effective ion--cavity coupling.
Further improvements are thus expected by encoding the logical qubit across more physical qubits, as in a planar microfabricated trap \cite{Cetina13}.
Maximum enhancement would be achieved by encoding not just one but $N/2$ excitations in a symmetrized $N$-ion state. 
The cooperative emission rate would then be $g\sqrt{\frac{N}{2}\left(\frac{N}{2}+1\right)}$, which scales with $N$ for large $N$, as observed in atomic ensembles~\cite{Baumann10,Baden14,Bohnet12}.
However, it remains an open question how to transfer quantum information between such 
states and single photons, as required for a quantum transducer~\cite{Lamata11}.

Finally, we emphasize two advantages of ions as qubits in these experiments: first, that the coupling strength of each ion to the cavity can be precisely controlled, and second, that a universal set of gate operations~\cite{Haeffner08} allows preparation of a range of states, from sub- to superradiant.
By tuning over this range, one could selectively turn off and on the coupling of logical qubits to the cavity.
This technique would provide a versatile tool for addressable read--write operations in a quantum register. 
}
%
%
\begin{acknowledgments}
We thank L. Lamata and F. Ong for helpful discussions and A. Stute for early contributions to the experiment design.
We gratefully acknowledge support from the Austrian Science Fund (FWF):  Project. Nos. F4003 and F4019, 
the European Commission via the Atomic QUantum TEchnologies (AQUTE) Integrating Project, 
and the Institut f\"ur Quanteninformation GmbH.
{ 
While preparing this manuscript, we learned of related work with two neutral atoms coupled to a cavity \cite{Reimann14}.}
\end{acknowledgments}


\section{Appendix}

\appendix

\section{System parameters}
Two $\Ca$ ions are confined in a linear Paul trap and coupled to an optical cavity.
The cavity decay rate is $\kappa = 2\pi \times 50$ kHz, and the atomic decay rate is $\gamma = 2\pi \times 11.5$~MHz, which is the sum of the decay channels from $\mathbb{P}$ to $ \mathbb D$ and from $\mathbb P$ to $ \mathbb S$, 
where the manifolds are defined as $\mathbb P \equiv 4^2P_{3/2} $, $ \mathbb D\equiv 3^2D_{5/2}$, and $\mathbb S\equiv 4^2S_{1/2}$.
The coupling strength of a single ion to the cavity mode on the $\mathbb {P - D}$ transition is $g^{\,}_{PD} =  2 \pi \times 1$~MHz. 
A Raman beam with Rabi frequency $\Omega$ is used to drive the $\mathbb {S - P}$ transitions.
The cavity parameters are described in further detail in Ref.~\cite{Casabone13Sup}. 

The three-level system $\mathbb S$-$\mathbb P$-$\mathbb D$ can be mapped onto an effective two-level system $\mathbb S$-$\mathbb D$ 
if a Raman resonance condition is met, i.e., when both Raman beam and cavity resonance have the same detuning from $\mathbb P$~\cite{Stute12a}.
During a cavity-mediated process, the electronic population of the ion is coherently transferred from a state in $\mathbb S$ to a state in $\mathbb D$, generating a cavity photon. For sufficiently large $\Delta$, negligible population is transferred to $\mathbb P$.
The rates of the effective two-level system are {
$g =\frac{\xi_{SD} \, \Omega \, g^{\,}_{PD}  }{2 \Delta}$} and $\gamma_{\text{eff}}=\gamma \left(\frac{\Omega}{2 \Delta}\right)^2$.
Here, $\Delta\sim 400$~MHz {
and $\xi_{SD}$ is a geometric factor that takes into account both the projection of the vacuum-mode polarization onto the atomic dipole moment and the Clebsch-Gordon coefficients of the $\mathbb {S - P}$ and $\mathbb {D - P}$ transitions~\cite{Stute12a}.}

Ten individual Raman transitions between  $\mathbb S$ and $\mathbb D$  
can be identified when all Zeeman sublevels are considered. 
A magnetic field of $B=4.5$~G, orthogonal to both the cavity axis and the wavevector of the Raman beam, lifts the degeneracy of the Zeeman sublevels such that each transition can be individually addressed~\cite{Stute12a}.
The strength of the magnetic field is determined via spectroscopy of the  $\mathbb S-\mathbb D$ transitions.

In the main text, two experiments are presented. 
In the first experiment,  we examine the probability to generate a photon as a function of the phase of the two-ion entangled state.
To perform the experiment, a Raman beam with Rabi frequency  {
$\Omega=19$~MHz} drives the $\ket S \equiv \ket{\mathbb S, m_j=-1/2}$ to $\ket D \equiv \ket{\mathbb D, m_j=-1/2}$ transition.
{
For $\Omega=19$~MHz, the rates of the effective two-level system are $\gamma_{\text{eff}}=2\pi \times 6$~kHz and $g_{\text{eff}}=2\pi \times 18$~kHz.
The cavity decay rate $\kappa = 2\pi \times 50$~kHz is the fastest of the three, placing the system in the bad cavity regime.}
In the second experiment, we use a superradiant state
to enhance the performance of a cavity-based quantum interface.
In this case, a bichromatic Raman beam with Rabi frequencies {
$19$ and $9.5$~MHz} drives the $\ket{S}$ to $ \ket{D}$ and   $\ket {S'} \equiv \ket{\mathbb S, m_j=1/2}$ to $\ket{D}$ transitions.
{
These transitions do not have equal transition probabilities and additionally, the orthogonally polarized photons couple differently to the cavity because of the orientation of the cavity with respect to the magnetic field~\cite{Stute13}. 
By choosing the Rabi frequency for the $\ket{S'}$ to $\ket{D}$ transition to have twice the value of the $\ket{S}$ to $\ket{D}$ transition, these differences are balanced and both transitions are driven with equal strength. }
In both experiments, the Rabi frequencies are first determined experimentally via Stark-shift measurements with an uncertainty on the order of {
10\%}.
Next, in simulations of single-photon generation, we adjust the Rabi frequencies within the experimental uncertainty range and find values for which the temporal photon shapes have the best agreement with data.

\section{Relative Raman phase}
In the first experiment, the part of the Hamiltonian that describes the interactions of the Raman laser and the cavity with the ion is 
\begin{align}
H_\text{int}  =  & g^{\,}_{PD} \, \big (\sigma_{PD}^{(1)} - \sigma^{(2)}_{PD}  \big )a^\dagger +  \nonumber  \\
& \Omega \, \big( e^{i\phi_{R_1}}\sigma^{(1)}_{SP} + e^{i\phi_{R_2}}\sigma^{(2)}_{SP}  \big)  + \text{h.c.},
\label{H_int} 
\tag{sm 1}
\end{align}
where $\sigma^{(i)}_{PD} \equiv \ket{D}\bra{P}$, $\sigma^{(i)}_{SP} \equiv \ket{P}\bra{S}$, $a^{\dagger}$ is the photon creation operator, and 
$\phi_{R_i}$ is the optical phase of the Raman beam when interacting with the $i$th ion.
Here, the rotating wave approximation has been used and an appropriate transformation to the interaction picture has been applied such that the Hamiltonian is time-independent. 
In this model, both ions are coupled to the cavity with the same strength, and the minus sign between the first and the second terms of Eq.~(\ref{H_int}) accounts for the fact that in our cavity system the two ions are located in adjacent antinodes~\cite{Casabone13}.

When the Raman resonance condition is met, Eq.~(\ref{H_int}) can be rewritten as Eq.~(1), identifying $\zeta = (\phi_{R_1}  - \phi_{R_2})$ and $\sigma_-=\ket{D}\bra{S}$.
The relative phase $\zeta$  is given by $\zeta=2 \pi\, d \, \sin\theta   / \lambda$, where $d$ is the ions' separation, $\theta \approx 45^\circ$ is the angle between trap axis and Raman beam, and $\lambda = 393$~nm is the wavelength of the Raman beam.
%

\section{Initial state preparation}

To generate $\ket{\Psi(\phi)} = \left(\ket{S}\ket{D}  + e^{i\phi} \ket{D}\ket{S}\right)/\sqrt{2}$, we first produce a maximally entangled state  {
$\ket{\Phi}= \big (\ket S \ket S +i\ket D \ket D\big )/\sqrt{2}$} by means of a M{\o}lmer--S{\o}rensen gate{
-operation}~\cite{Sorensen99}.
To perform the gate, we off-resonantly drive the blue and red sidebands of the axial center-of-mass motion of the $\ket S \leftrightarrow \ket D$ transition with a detuning $\delta$.
The ions are initialized in $\ket S\ket S$.  After a time $T=1/\delta=55~\mu$s, with a detuning $\delta=18.2$~kHz, the two ions are prepared in the entangled state {
$\ket \Phi$} (see Fig. (1b)).
For comparison, the coherence time for information stored in the $\mathbb {S-D}$ qubit is 475~$\mu$s.

We calculate the fidelity {
$F_{\Phi}$} of the experimental state with respect to  {
$\ket {\Phi}$ } in the following way ~\cite{Benhelm08}.
After {
$\ket {\Phi}$ }is created, we apply an `analysis' $\pi/2$ pulse on the $\ket S \leftrightarrow  \ket D$ transition with a variable phase with respect to the previous entangling pulse.
Subsequently, the parity operator $P=p_{SS} + p_{DD} - p_{SD,DS}$ is calculated from fluorescence measurements of the ion populations, 
where $p_{SS}$ and $p_{DD}$ are the probabilities to find both ions in $\ket S \ket S$ and $\ket D \ket D$, respectively,
and $p_{SD,DS}$ is the probability to find one ion in $\ket S$ and the other in $\ket D$.
Fig. (1c) shows the parity $P$ as function of the phase of the analysis pulse.
If $A$ is the amplitude of the parity oscillation, then the fidelity  {
$F_{\Phi}$} is bound from above via {
$F_{\Phi}\ge A$}. 
From a fit to the data of Fig (2a), we calculate that {
$\ket {\Phi}$} is created with a fidelity of at least 95(2)\%. 

After the state {
$\ket {\Phi}$} is generated, a $\pi/2$-pulse on the $\ket S \leftrightarrow  \ket D$ transition rotates the state to $(\ket S \ket D+\ket D \ket S)/\sqrt{2}$, identified in Fig. (1c).
%
%
Finally, to convert $(\ket S \ket D +\ket D \ket S)/\sqrt{2}$ to $\ket{\Psi(\phi)}$, we perform a single-ion rotation, introducing AC-Stark shifts to one ion using the addressing beam~\cite{schindler13}.
The phase $\phi$ of $\ket{\Psi(\phi)}$ is proportional to the duration $\tau$ of the Stark-shift pulse,
where the proportionality constant depends on the Rabi frequency of the addressing beam, $\Omega_{\text{AC}}$, and the detuning of the laser from the $\ket S \leftrightarrow \ket D$ transition, $\delta_\text{AC}$.
We choose $\delta_\text{AC}=10$~MHz and $\Omega_\text{AC}=8.6$~MHz for a rotation that has a period of $5.3~ \mu$s. 

\begin{figure}
	\includegraphics[width=0.475\textwidth]{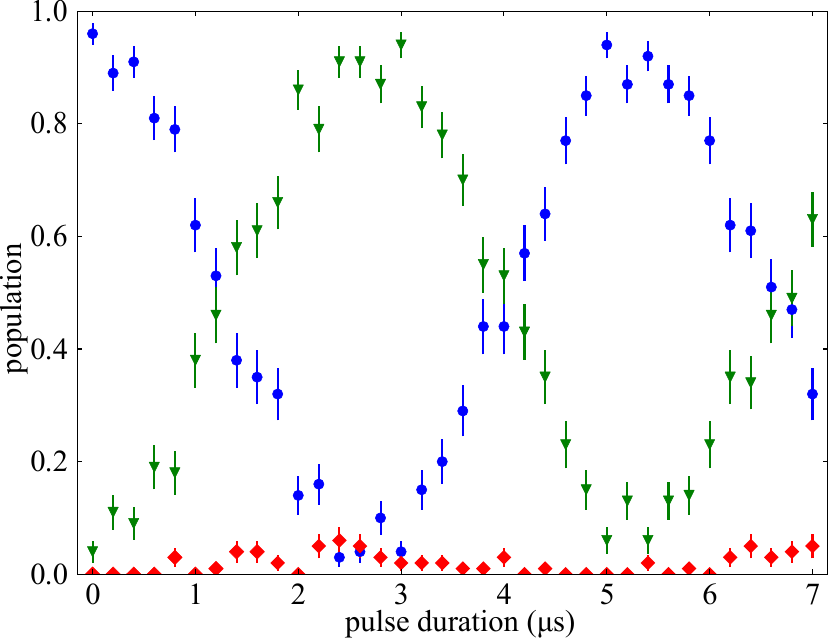}
	\caption{Populations of the states $\ket{S}\ket{S}$ (red diamonds), $\ket{D}\ket{D}$ (blue circles) and $\ket{S}\ket{D}$ (green triangles) as function of the duration of the AC-Stark shift pulse. 
	After $2.6~\mu$s, state $\ket S \ket D$ is generated with a fidelity of 91(4)\%.
	Error bars represent projection noise.} \label{fig4}
\end{figure}

The implementation of the Stark-shift gate is demonstrated via the generation of the state $\ket S \ket D$.
After optical pumping of both ions to $\ket S \ket S$, we apply a $\pi/2$ rotation on the $\ket S \leftrightarrow  \ket D$ transition using the global beam.
Next, the Stark-shift gate is applied to one ion for a time $\tau$.
Subsequently, another global $\pi/2$ rotation on the $\ket S \leftrightarrow  \ket D$ transition is applied with the same phase as the first $\pi/2$ rotation.
Finally, $\ket S \ket S$, $\ket D \ket D$ and $\ket S \ket D$ populations are extracted via fluorescence detection.
The results are shown in Fig (\ref{fig4}) as a function of $\tau$. 
After $2.6~\mu$s, {
the ions are in a state with a fidelity of 91(4)\% with respect to $\ket S \ket D$.}

{
There are at least two other methods by which one could tune the phase $\phi$ in the experiment.  
First, the angle of the Raman beam could be changed.  
Second, the ion--ion separation could be changed by means of the voltages that determine the trap potential.  
Both methods would shift the relative phase seen by each ion.  
In initial experiments, we used the second method; however, when the ion--ion separation is adjusted to correspond to a desired phase, both ions must also remain equally and near-maximally coupled to the cavity \cite{Casabone13}, and it is not straightforward to satisfy both conditions simultaneously.
In practice, we found the Stark-shift gate described above to be the most precise and reproducible approach.  }
 
To generate the single ion cases $\ket{\psi_1}$ and $\ket{\psi_2}$, we use the addressing beam.
In this case, the frequency of the addressing beam is set to drive the $\ket S \leftrightarrow \ket{D'}\equiv \ket{\mathbb D, m_j=3/2}$ transition on resonance.
As the addressing beam interacts with the second ion, a $\pi$-pulse transfers the state $\ket S \ket S$ to $\ket{\psi_1} =  \ket S \ket{D'}$.
To generate $\ket{\psi_2}  = \ket{D'} \ket {S}$, we subsequently apply a $\pi$-rotation on the $\ket S \leftrightarrow \ket{D'}$ transition to both ions, such that $\ket S \ket{D'}$ is rotated to $\ket {D'} \ket {S}$.
The single-ion cases are prepared with a fidelity of 95(3)\%.  

\section{Two-ion crystal as a single--photon source}
We have previously demonstrated that one ion in $\mathbb S$ produces a single photon when a Raman transition between  $\mathbb S$ and  $\mathbb D$ is driven~\cite{Barros09}.
In the experiments presented in the main text, we consider two ions in the entangled state $\ket{\Psi(\phi)}$, in which the probability to find one ion in $\ket S$ is one.
When a Raman transition is driven between $\ket S$ and $\ket D$, the entangled state $\ket{\Psi(\phi)}$ is transferred to $\ket D \ket D$ and a single photon is expected.
However, imperfect preparation of $\ket{\Psi(\phi)}$ leaves some population in $\ket S \ket S$, resulting in the generation of two photons.

In order to estimate the number of two-photon detection events, we consider detector dark counts and imperfect preparation of the ions' state.
The following four events are relevant and contribute to two-photon detections:
\begin{enumerate}
 	\item State $\ket S \ket S$ is generated; two photons are produced and detected.
	 \item State $\ket S \ket S$ is generated; two photons are produced, one is lost and the other is detected together with a dark count.
	 \item State $\ket{\Psi (\phi)}$ is generated; one photon is produced and is detected together with a dark count.
	 \item Two darks count are detected.
\end{enumerate}

State tomography reveals that in 3(2)\% of attempts to generate $\ket{\Psi(\phi)}$, the state $\ket S \ket S$ is prepared instead.
The probability to detect one photon during the $55~\mu$s duration of the Raman process is 5.4(3)\%, which is mainly limited by cavity absorption and detector efficiencies~\cite{Stute12a}. 
Detector dark count rates are 3.2(1)~s$^{-1}$ and 3.8(1)~s$^{-1}$ for the two avalanche photodiodes.
With these values, we expect one two-photon event in $8.2(8) \times 10^3$ attempts to generate a single photon. 

To measure two-photon events, we generate  $\big (\ket S \ket D + \ket D \ket S \big ) / \sqrt{2}$ and $\big (\ket{S'} \ket D + \ket D \ket{S'} \big ) / \sqrt{2}$ and drive a cavity-mediated Raman transition such that a horizontally or a vertically polarized photon is generated.
Photons leaving the cavity cross a half-wave plate aligned such that 50\% of the light is reflected and 50\% transmitted by a polarizing  beam splitter.
Photons are detected by the two avalanche photodiodes at each beam-splitter output and the second-order correlation function $g^{(2)}(0)$ is calculated.
After 223,106 attempts to generate photons, 28 two--photon events were measured, and 27(3) two--photon events were expected from the considerations above.
The observed number of two-photon detection events are thus consistent with single-photon generation.

\section{Process fidelity}
Tomography of the state-mapping process consists of state tomography of the photonic output qubit for four orthogonal input states. 
Measurements in the three bases of horizontal/vertical, diagonal/antidiagonal and right/left circular polarization constitute state tomography of the photonic qubit~\cite{Nielsen2000}. 
Each basis is measured a second time with the APDs swapped by rotating the $\lambda/2$- and $\lambda/4$-waveplates. 
Analysis is done with the sum of the two measurements to compensate for the different detection efficiencies of the two APDs. 

Process matrices $\chi_{ij}$ are reconstructed using a maximum-likelihood method. 
The process fidelity $\chi_{00}$ is given by the overlap of the reconstructed process matrix with the target process (i.e., the identity operation). 
Uncertainties in the process fidelities are given as one standard deviation, derived from non-parametric bootstrapping assuming a multinomial distribution~\cite{Efron93}. 


\section{Simulations}
Numerical simulations are based on the Quantum Optics and Computation Toolbox for MATLAB~\cite{Tan99} via integration of the master equation. 
We simulate two $\Ca$ ions interacting with an optical cavity and a Raman beam.
For each ion, we consider six levels:  $\ket{S}$, $\ket{S'}$, $\ket{D}$, $\ket{D'}$, $\ket {\mathbb P, m_j=-1/2}$ and $\ket {\mathbb P, m_j=1/2}$.
For the optical cavity, we consider two orthogonal modes $a $ and $ b$ with the Fock state basis truncated at 2 for each mode.

Additional input parameters for the simulations are the cavity parameters $g, \kappa$, and $\gamma$; the magnetic field amplitude $B$, the Rabi frequency $\Omega$ of the Raman laser, the Raman laser linewidth, and the output path losses.
The laser linewidth, atomic decay, and cavity decay are introduced in the Lindblad form.
The Raman laser linewidth is set to the measured value of $30$~kHz.

For the simulation of the first experiment, the initial density matrix $\rho_0$ is assigned {
5\%} of populations equally distributed between $\ket S\ket S$ and $\ket D\ket D$, and the coherence terms between $\ket S\ket S$ and $\ket D\ket D$ are set to zero, consistent with measurements.
The rest of the population is distributed between $\ket S\ket D$ and $\ket D\ket S$, preserving the coherences such that $\rho_0$ has an overlap of {
95\%} with $\ket {\Psi (\varphi)}$.
In the case of the second experiment, the superposition state
\[
	\ket{\alpha,\beta} \equiv \cos \alpha \ket S + e^{i\beta} \sin \alpha \ket {S'},
\]
is introduced 
via an operator $\hat M$ that performs the mapping 
\[
\ket S \to \cos \alpha \ket S + e^{i\beta} \sin \alpha \ket {S'}
\]
for each ion.
This operator $\hat M$ is applied to $\rho_0$.

From the integration of the master equation up to a time $t$, we obtain the time-dependent density matrix $\rho(t)$.
The mean photon numbers of the cavity modes are calculated via the expectation values $\langle a^\dagger a (t)\rangle$ and $\langle b^\dagger b (t) \rangle$.
Contributions of the detector dark counts are added to the mean photon number.
Errors in the generation of the superposition state and magnetic field fluctuations are introduced by scaling the off-diagonal terms of $\rho(t)$ by a factor of 0.96 and by the exponential $e^{(2t/\tau)^2}$ respectively, where $\tau=190~\mu$s is the coherence time of the qubit stored in $\ket S$ and $\ket{S'}$.
{
Finally, the coupling of one of the ions to the cavity mode is reduced to 90\% of it maximum value. 
This reduction is based on measured drifts over the course of the experiment.}

Fig. (2c) shows the simulated and experimental temporal photon shapes as function of detection time.
In order to have good agreement between the experimental data and simulations, we adjusted the Rabi frequency $\Omega$ within the experimental uncertainty range (see the ``System Parameters" section).

In the main text, we note that scattering and imperfect state preparation reduce the photon generation probability of the entangled state during the first 6~$\mu$s of the Raman process.
In order to quantify this effect, we simulate the temporal photon shape as a function of detection time for the superradiant state, for the case of perfect state initialization and no scattering from  $\mathbb P$ to $\mathbb S$ and $\mathbb D$.
Comparing the area under this curve to that from the simulation in Fig. (2c), which takes both imperfect state initialization and scattering into account, we extract a reduction of 9.1\%.

\bibliography{bibliography,misc_arxiv}

\begin{thebibliography}{39}%
\makeatletter
\providecommand \@ifxundefined [1]{%
 \@ifx{#1\undefined}
}%
\providecommand \@ifnum [1]{%
 \ifnum #1\expandafter \@firstoftwo
 \else \expandafter \@secondoftwo
 \fi
}%
\providecommand \@ifx [1]{%
 \ifx #1\expandafter \@firstoftwo
 \else \expandafter \@secondoftwo
 \fi
}%
\providecommand \natexlab [1]{#1}%
\providecommand \enquote  [1]{``#1''}%
\providecommand \bibnamefont  [1]{#1}%
\providecommand \bibfnamefont [1]{#1}%
\providecommand \citenamefont [1]{#1}%
\providecommand \href@noop [0]{\@secondoftwo}%
\providecommand \href [0]{\begingroup \@sanitize@url \@href}%
\providecommand \@href[1]{\@@startlink{#1}\@@href}%
\providecommand \@@href[1]{\endgroup#1\@@endlink}%
\providecommand \@sanitize@url [0]{\catcode `\\12\catcode `\$12\catcode
  `\&12\catcode `\#12\catcode `\^12\catcode `\_12\catcode `\%12\relax}%
\providecommand \@@startlink[1]{}%
\providecommand \@@endlink[0]{}%
\providecommand \url  [0]{\begingroup\@sanitize@url \@url }%
\providecommand \@url [1]{\endgroup\@href {#1}{\urlprefix }}%
\providecommand \urlprefix  [0]{URL }%
\providecommand \Eprint [0]{\href }%
\providecommand \doibase [0]{http://dx.doi.org/}%
\providecommand \selectlanguage [0]{\@gobble}%
\providecommand \bibinfo  [0]{\@secondoftwo}%
\providecommand \bibfield  [0]{\@secondoftwo}%
\providecommand \translation [1]{[#1]}%
\providecommand \BibitemOpen [0]{}%
\providecommand \bibitemStop [0]{}%
\providecommand \bibitemNoStop [0]{.\EOS\space}%
\providecommand \EOS [0]{\spacefactor3000\relax}%
\providecommand \BibitemShut  [1]{\csname bibitem#1\endcsname}%
\let\auto@bib@innerbib\@empty
\bibitem [{\citenamefont {Dicke}(1954)}]{Dicke54}%
  \BibitemOpen
  \bibfield  {author} {\bibinfo {author} {\bibfnamefont {R.~H.}\ \bibnamefont
  {Dicke}},\ }\href {\doibase 10.1103/PhysRev.93.99} {\bibfield  {journal}
  {\bibinfo  {journal} {Phys. Rev.}\ }\textbf {\bibinfo {volume} {93}},\
  \bibinfo {pages} {99} (\bibinfo {year} {1954})}\BibitemShut {NoStop}%
\bibitem [{\citenamefont {Gross}\ and\ \citenamefont
  {Haroche}(1982)}]{Gross82}%
  \BibitemOpen
  \bibfield  {author} {\bibinfo {author} {\bibfnamefont {M.}~\bibnamefont
  {Gross}}\ and\ \bibinfo {author} {\bibfnamefont {S.}~\bibnamefont
  {Haroche}},\ }\href@noop {} {\bibfield  {journal} {\bibinfo  {journal} {Phys.
  Rep.}\ }\textbf {\bibinfo {volume} {93}},\ \bibinfo {pages} {301} (\bibinfo
  {year} {1982})}\BibitemShut {NoStop}%
\bibitem [{\citenamefont {Garraway}(2011)}]{Garraway11}%
  \BibitemOpen
  \bibfield  {author} {\bibinfo {author} {\bibfnamefont {B.~M.}\ \bibnamefont
  {Garraway}},\ }\href@noop {} {\bibfield  {journal} {\bibinfo  {journal}
  {Phil. Trans. R. Soc. A}\ }\textbf {\bibinfo {volume} {369}},\ \bibinfo
  {pages} {1137} (\bibinfo {year} {2011})}\BibitemShut {NoStop}%
\bibitem [{\citenamefont {Baumann}\ \emph {et~al.}(2010)\citenamefont
  {Baumann}, \citenamefont {Guerlin}, \citenamefont {Brennecke},\ and\
  \citenamefont {Esslinger}}]{Baumann10}%
  \BibitemOpen
  \bibfield  {author} {\bibinfo {author} {\bibfnamefont {K.}~\bibnamefont
  {Baumann}}, \bibinfo {author} {\bibfnamefont {C.}~\bibnamefont {Guerlin}},
  \bibinfo {author} {\bibfnamefont {F.}~\bibnamefont {Brennecke}}, \ and\
  \bibinfo {author} {\bibfnamefont {T.}~\bibnamefont {Esslinger}},\ }\href@noop
  {} {\bibfield  {journal} {\bibinfo  {journal} {Nature}\ }\textbf {\bibinfo
  {volume} {464}},\ \bibinfo {pages} {1301} (\bibinfo {year}
  {2010})}\BibitemShut {NoStop}%
\bibitem [{\citenamefont {Baden}\ \emph {et~al.}(2014)\citenamefont {Baden},
  \citenamefont {Arnold}, \citenamefont {Grimsmo}, \citenamefont {Parkins},\
  and\ \citenamefont {Barrett}}]{Baden14}%
  \BibitemOpen
  \bibfield  {author} {\bibinfo {author} {\bibfnamefont {M.~P.}\ \bibnamefont
  {Baden}}, \bibinfo {author} {\bibfnamefont {K.~J.}\ \bibnamefont {Arnold}},
  \bibinfo {author} {\bibfnamefont {A.~L.}\ \bibnamefont {Grimsmo}}, \bibinfo
  {author} {\bibfnamefont {S.}~\bibnamefont {Parkins}}, \ and\ \bibinfo
  {author} {\bibfnamefont {M.~D.}\ \bibnamefont {Barrett}},\ }\href@noop {} {\
  (\bibinfo {year} {2014})},\ \Eprint {http://arxiv.org/abs/1404.0512
  [quant-ph]} {arXiv:1404.0512 [quant-ph]} \BibitemShut {NoStop}%
\bibitem [{\citenamefont {Bohnet}\ \emph {et~al.}(2012)\citenamefont {Bohnet},
  \citenamefont {Chen}, \citenamefont {Weiner}, \citenamefont {Meiser},
  \citenamefont {Holland},\ and\ \citenamefont {Thompson}}]{Bohnet12}%
  \BibitemOpen
  \bibfield  {author} {\bibinfo {author} {\bibfnamefont {J.~G.}\ \bibnamefont
  {Bohnet}}, \bibinfo {author} {\bibfnamefont {Z.}~\bibnamefont {Chen}},
  \bibinfo {author} {\bibfnamefont {J.~M.}\ \bibnamefont {Weiner}}, \bibinfo
  {author} {\bibfnamefont {D.}~\bibnamefont {Meiser}}, \bibinfo {author}
  {\bibfnamefont {M.~J.}\ \bibnamefont {Holland}}, \ and\ \bibinfo {author}
  {\bibfnamefont {J.~K.}\ \bibnamefont {Thompson}},\ }\href@noop {} {\bibfield
  {journal} {\bibinfo  {journal} {Nature}\ }\textbf {\bibinfo {volume} {484}},\
  \bibinfo {pages} {78} (\bibinfo {year} {2012})}\BibitemShut {NoStop}%
\bibitem [{\citenamefont {Lukin}\ \emph {et~al.}(2001)\citenamefont {Lukin},
  \citenamefont {Fleischhauer}, \citenamefont {Cote}, \citenamefont {Duan},
  \citenamefont {Jaksch}, \citenamefont {Cirac},\ and\ \citenamefont
  {Zoller}}]{Lukin01}%
  \BibitemOpen
  \bibfield  {author} {\bibinfo {author} {\bibfnamefont {M.~D.}\ \bibnamefont
  {Lukin}}, \bibinfo {author} {\bibfnamefont {M.}~\bibnamefont {Fleischhauer}},
  \bibinfo {author} {\bibfnamefont {R.}~\bibnamefont {Cote}}, \bibinfo {author}
  {\bibfnamefont {L.~M.}\ \bibnamefont {Duan}}, \bibinfo {author}
  {\bibfnamefont {D.}~\bibnamefont {Jaksch}}, \bibinfo {author} {\bibfnamefont
  {J.~I.}\ \bibnamefont {Cirac}}, \ and\ \bibinfo {author} {\bibfnamefont
  {P.}~\bibnamefont {Zoller}},\ }\href {\doibase 10.1103/PhysRevLett.87.037901}
  {\bibfield  {journal} {\bibinfo  {journal} {Phys. Rev. Lett.}\ }\textbf
  {\bibinfo {volume} {87}},\ \bibinfo {pages} {037901} (\bibinfo {year}
  {2001})}\BibitemShut {NoStop}%
\bibitem [{\citenamefont {Meir}\ \emph {et~al.}(2013)\citenamefont {Meir},
  \citenamefont {Schwartz}, \citenamefont {Shahmoon}, \citenamefont {Oron},\
  and\ \citenamefont {Ozeri}}]{Meir13}%
  \BibitemOpen
  \bibfield  {author} {\bibinfo {author} {\bibfnamefont {Z.}~\bibnamefont
  {Meir}}, \bibinfo {author} {\bibfnamefont {O.}~\bibnamefont {Schwartz}},
  \bibinfo {author} {\bibfnamefont {E.}~\bibnamefont {Shahmoon}}, \bibinfo
  {author} {\bibfnamefont {D.}~\bibnamefont {Oron}}, \ and\ \bibinfo {author}
  {\bibfnamefont {R.}~\bibnamefont {Ozeri}},\ }\href@noop {} {\  (\bibinfo
  {year} {2013})},\ \Eprint {http://arxiv.org/abs/1312.5933 [quant-ph]}
  {arXiv:1312.5933 [quant-ph]} \BibitemShut {NoStop}%
\bibitem [{\citenamefont {DeVoe}\ and\ \citenamefont
  {Brewer}(1996)}]{DeVoe1996}%
  \BibitemOpen
  \bibfield  {author} {\bibinfo {author} {\bibfnamefont {R.~G.}\ \bibnamefont
  {DeVoe}}\ and\ \bibinfo {author} {\bibfnamefont {R.~G.}\ \bibnamefont
  {Brewer}},\ }\href {\doibase 10.1103/PhysRevLett.76.2049} {\bibfield
  {journal} {\bibinfo  {journal} {Phys. Rev. Lett.}\ }\textbf {\bibinfo
  {volume} {76}},\ \bibinfo {pages} {2049} (\bibinfo {year}
  {1996})}\BibitemShut {NoStop}%
\bibitem [{\citenamefont {Kimble}(2008)}]{Kimble08a}%
  \BibitemOpen
  \bibfield  {author} {\bibinfo {author} {\bibfnamefont {H.~J.}\ \bibnamefont
  {Kimble}},\ }\href@noop {} {\bibfield  {journal} {\bibinfo  {journal}
  {Nature}\ }\textbf {\bibinfo {volume} {453}},\ \bibinfo {pages} {1023}
  (\bibinfo {year} {2008})}\BibitemShut {NoStop}%
\bibitem [{\citenamefont {Duan}\ and\ \citenamefont {Monroe}(2010)}]{Duan10}%
  \BibitemOpen
  \bibfield  {author} {\bibinfo {author} {\bibfnamefont {L.-M.}\ \bibnamefont
  {Duan}}\ and\ \bibinfo {author} {\bibfnamefont {C.}~\bibnamefont {Monroe}},\
  }\href {\doibase 10.1103/RevModPhys.82.1209} {\bibfield  {journal} {\bibinfo
  {journal} {Rev. Mod. Phys.}\ }\textbf {\bibinfo {volume} {82}},\ \bibinfo
  {pages} {1209} (\bibinfo {year} {2010})}\BibitemShut {NoStop}%
\bibitem [{\citenamefont {Duan}\ \emph {et~al.}(2001)\citenamefont {Duan},
  \citenamefont {Lukin}, \citenamefont {Cirac},\ and\ \citenamefont
  {Zoller}}]{Duan01}%
  \BibitemOpen
  \bibfield  {author} {\bibinfo {author} {\bibfnamefont {L.-M.}\ \bibnamefont
  {Duan}}, \bibinfo {author} {\bibfnamefont {M.~D.}\ \bibnamefont {Lukin}},
  \bibinfo {author} {\bibfnamefont {J.~I.}\ \bibnamefont {Cirac}}, \ and\
  \bibinfo {author} {\bibfnamefont {P.}~\bibnamefont {Zoller}},\ }\href@noop {}
  {\bibfield  {journal} {\bibinfo  {journal} {Nature}\ }\textbf {\bibinfo
  {volume} {414}},\ \bibinfo {pages} {413} (\bibinfo {year}
  {2001})}\BibitemShut {NoStop}%
\bibitem [{\citenamefont {de~Oliveira}\ \emph {et~al.}(2014)\citenamefont
  {de~Oliveira}, \citenamefont {Mendes}, \citenamefont {Martins}, \citenamefont
  {Saldanha}, \citenamefont {Tabosa},\ and\ \citenamefont
  {Felinto}}]{Oliveira14}%
  \BibitemOpen
  \bibfield  {author} {\bibinfo {author} {\bibfnamefont {R.~A.}\ \bibnamefont
  {de~Oliveira}}, \bibinfo {author} {\bibfnamefont {M.~S.}\ \bibnamefont
  {Mendes}}, \bibinfo {author} {\bibfnamefont {W.~S.}\ \bibnamefont {Martins}},
  \bibinfo {author} {\bibfnamefont {P.~L.}\ \bibnamefont {Saldanha}}, \bibinfo
  {author} {\bibfnamefont {J.~W.~R.}\ \bibnamefont {Tabosa}}, \ and\ \bibinfo
  {author} {\bibfnamefont {D.}~\bibnamefont {Felinto}},\ }\href {\doibase
  10.1103/PhysRevA.90.023848} {\bibfield  {journal} {\bibinfo  {journal} {Phys.
  Rev. A}\ }\textbf {\bibinfo {volume} {90}},\ \bibinfo {pages} {023848}
  (\bibinfo {year} {2014})}\BibitemShut {NoStop}%
\bibitem [{\citenamefont {Cirac}\ \emph {et~al.}(1997)\citenamefont {Cirac},
  \citenamefont {Zoller}, \citenamefont {Kimble},\ and\ \citenamefont
  {Mabuchi}}]{Cirac97}%
  \BibitemOpen
  \bibfield  {author} {\bibinfo {author} {\bibfnamefont {J.~I.}\ \bibnamefont
  {Cirac}}, \bibinfo {author} {\bibfnamefont {P.}~\bibnamefont {Zoller}},
  \bibinfo {author} {\bibfnamefont {H.~J.}\ \bibnamefont {Kimble}}, \ and\
  \bibinfo {author} {\bibfnamefont {H.}~\bibnamefont {Mabuchi}},\ }\href
  {\doibase 10.1103/PhysRevLett.78.3221} {\bibfield  {journal} {\bibinfo
  {journal} {Phys. Rev. Lett.}\ }\textbf {\bibinfo {volume} {78}},\ \bibinfo
  {pages} {3221} (\bibinfo {year} {1997})}\BibitemShut {NoStop}%
\bibitem [{\citenamefont {Lamata}\ \emph {et~al.}(2011)\citenamefont {Lamata},
  \citenamefont {Leibrandt}, \citenamefont {Chuang}, \citenamefont {Cirac},
  \citenamefont {Lukin}, \citenamefont {Vuleti\ifmmode~\acute{c}\else
  \'{c}\fi{}},\ and\ \citenamefont {Yelin}}]{Lamata11}%
  \BibitemOpen
  \bibfield  {author} {\bibinfo {author} {\bibfnamefont {L.}~\bibnamefont
  {Lamata}}, \bibinfo {author} {\bibfnamefont {D.~R.}\ \bibnamefont
  {Leibrandt}}, \bibinfo {author} {\bibfnamefont {I.~L.}\ \bibnamefont
  {Chuang}}, \bibinfo {author} {\bibfnamefont {J.~I.}\ \bibnamefont {Cirac}},
  \bibinfo {author} {\bibfnamefont {M.~D.}\ \bibnamefont {Lukin}}, \bibinfo
  {author} {\bibfnamefont {V.}~\bibnamefont {Vuleti\ifmmode~\acute{c}\else
  \'{c}\fi{}}}, \ and\ \bibinfo {author} {\bibfnamefont {S.~F.}\ \bibnamefont
  {Yelin}},\ }\href {\doibase 10.1103/PhysRevLett.107.030501} {\bibfield
  {journal} {\bibinfo  {journal} {Phys. Rev. Lett.}\ }\textbf {\bibinfo
  {volume} {107}},\ \bibinfo {pages} {030501} (\bibinfo {year}
  {2011})}\BibitemShut {NoStop}%
\bibitem [{\citenamefont {Plenio}\ \emph {et~al.}(1999)\citenamefont {Plenio},
  \citenamefont {Huelga}, \citenamefont {Beige},\ and\ \citenamefont
  {Knight}}]{Plenio99}%
  \BibitemOpen
  \bibfield  {author} {\bibinfo {author} {\bibfnamefont {M.~B.}\ \bibnamefont
  {Plenio}}, \bibinfo {author} {\bibfnamefont {S.~F.}\ \bibnamefont {Huelga}},
  \bibinfo {author} {\bibfnamefont {A.}~\bibnamefont {Beige}}, \ and\ \bibinfo
  {author} {\bibfnamefont {P.~L.}\ \bibnamefont {Knight}},\ }\href {\doibase
  10.1103/PhysRevA.59.2468} {\bibfield  {journal} {\bibinfo  {journal} {Phys.
  Rev. A}\ }\textbf {\bibinfo {volume} {59}},\ \bibinfo {pages} {2468}
  (\bibinfo {year} {1999})}\BibitemShut {NoStop}%
\bibitem [{\citenamefont {Beige}\ \emph {et~al.}(2000)\citenamefont {Beige},
  \citenamefont {Braun}, \citenamefont {Tregenna},\ and\ \citenamefont
  {Knight}}]{Beige00}%
  \BibitemOpen
  \bibfield  {author} {\bibinfo {author} {\bibfnamefont {A.}~\bibnamefont
  {Beige}}, \bibinfo {author} {\bibfnamefont {D.}~\bibnamefont {Braun}},
  \bibinfo {author} {\bibfnamefont {B.}~\bibnamefont {Tregenna}}, \ and\
  \bibinfo {author} {\bibfnamefont {P.~L.}\ \bibnamefont {Knight}},\ }\href
  {\doibase 10.1103/PhysRevLett.85.1762} {\bibfield  {journal} {\bibinfo
  {journal} {Phys. Rev. Lett.}\ }\textbf {\bibinfo {volume} {85}},\ \bibinfo
  {pages} {1762} (\bibinfo {year} {2000})}\BibitemShut {NoStop}%
\bibitem [{\citenamefont {Lidar}\ and\ \citenamefont
  {Birgitta~Whaley}(2003)}]{Lidar03}%
  \BibitemOpen
  \bibfield  {author} {\bibinfo {author} {\bibfnamefont {D.~A.}\ \bibnamefont
  {Lidar}}\ and\ \bibinfo {author} {\bibfnamefont {K.}~\bibnamefont
  {Birgitta~Whaley}},\ }in\ \href {\doibase 10.1007/3-540-44874-8_5} {\emph
  {\bibinfo {booktitle} {Irreversible Quantum Dynamics}}},\ \bibinfo {series}
  {Lecture Notes in Physics}, Vol.\ \bibinfo {volume} {622},\ \bibinfo {editor}
  {edited by\ \bibinfo {editor} {\bibfnamefont {F.}~\bibnamefont {Benatti}}\
  and\ \bibinfo {editor} {\bibfnamefont {R.}~\bibnamefont {Floreanini}}}\
  (\bibinfo  {publisher} {Springer Berlin Heidelberg},\ \bibinfo {year}
  {2003})\ pp.\ \bibinfo {pages} {83--120}\BibitemShut {NoStop}%
\bibitem [{\citenamefont {Filipp}\ \emph {et~al.}(2011)\citenamefont {Filipp},
  \citenamefont {van Loo}, \citenamefont {Baur}, \citenamefont {Steffen},\ and\
  \citenamefont {Wallraff}}]{Filipp11}%
  \BibitemOpen
  \bibfield  {author} {\bibinfo {author} {\bibfnamefont {S.}~\bibnamefont
  {Filipp}}, \bibinfo {author} {\bibfnamefont {A.~F.}\ \bibnamefont {van Loo}},
  \bibinfo {author} {\bibfnamefont {M.}~\bibnamefont {Baur}}, \bibinfo {author}
  {\bibfnamefont {L.}~\bibnamefont {Steffen}}, \ and\ \bibinfo {author}
  {\bibfnamefont {A.}~\bibnamefont {Wallraff}},\ }\href {\doibase
  10.1103/PhysRevA.84.061805} {\bibfield  {journal} {\bibinfo  {journal} {Phys.
  Rev. A}\ }\textbf {\bibinfo {volume} {84}},\ \bibinfo {pages} {061805}
  (\bibinfo {year} {2011})}\BibitemShut {NoStop}%
\bibitem [{\citenamefont {Benhelm}\ \emph {et~al.}(2008)\citenamefont
  {Benhelm}, \citenamefont {Kirchmair}, \citenamefont {Roos},\ and\
  \citenamefont {Blatt}}]{Benhelm08}%
  \BibitemOpen
  \bibfield  {author} {\bibinfo {author} {\bibfnamefont {J.}~\bibnamefont
  {Benhelm}}, \bibinfo {author} {\bibfnamefont {G.}~\bibnamefont {Kirchmair}},
  \bibinfo {author} {\bibfnamefont {C.~F.}\ \bibnamefont {Roos}}, \ and\
  \bibinfo {author} {\bibfnamefont {R.}~\bibnamefont {Blatt}},\ }\href
  {http://dx.doi.org/10.1038/nphys961} {\bibfield  {journal} {\bibinfo
  {journal} {Nat. Phys.}\ }\textbf {\bibinfo {volume} {4}},\ \bibinfo {pages}
  {463} (\bibinfo {year} {2008})}\BibitemShut {NoStop}%
\bibitem [{\citenamefont {Tavis}\ and\ \citenamefont
  {Cummings}(1968)}]{Tavis68}%
  \BibitemOpen
  \bibfield  {author} {\bibinfo {author} {\bibfnamefont {M.}~\bibnamefont
  {Tavis}}\ and\ \bibinfo {author} {\bibfnamefont {F.~W.}\ \bibnamefont
  {Cummings}},\ }\href {\doibase 10.1103/PhysRev.170.379} {\bibfield  {journal}
  {\bibinfo  {journal} {Phys. Rev.}\ }\textbf {\bibinfo {volume} {170}},\
  \bibinfo {pages} {379} (\bibinfo {year} {1968})}\BibitemShut {NoStop}%
\bibitem [{SM()}]{SM}%
  \BibitemOpen
  \href@noop {} {\bibinfo  {journal} {See Appendix for details about the
  relative Raman phase, cavity coupling parameters, ion--ion entanglement, and
  simulations}\ }\BibitemShut {NoStop}%
\bibitem [{\citenamefont {Nielsen}\ and\ \citenamefont
  {Chuang}(2010)}]{Nielsen2000}%
  \BibitemOpen
\bibfield  {journal} {  }\bibfield  {author} {\bibinfo {author} {\bibfnamefont
  {M.~A.}\ \bibnamefont {Nielsen}}\ and\ \bibinfo {author} {\bibfnamefont
  {I.~L.}\ \bibnamefont {Chuang}},\ }\href
  {http://www.cambridge.org/gb/knowledge/isbn/item5731729/?site_locale=en_GB}
  {\emph {\bibinfo {title} {Quantum Computation and Quantum Information}}}\
  (\bibinfo  {publisher} {Cambridge University Press},\ \bibinfo {address}
  {Cambridge},\ \bibinfo {year} {2010})\BibitemShut {NoStop}%
\bibitem [{\citenamefont {Efron}\ and\ \citenamefont
  {Tibshirani}(1993)}]{Efron93}%
  \BibitemOpen
  \bibfield  {author} {\bibinfo {author} {\bibfnamefont {B.}~\bibnamefont
  {Efron}}\ and\ \bibinfo {author} {\bibfnamefont {R.}~\bibnamefont
  {Tibshirani}},\ }\href@noop {} {\emph {\bibinfo {title} {An introduction to
  the bootstrap}}}\ (\bibinfo  {publisher} {Chapman \& Hall},\ \bibinfo
  {address} {New York},\ \bibinfo {year} {1993})\BibitemShut {NoStop}%
\bibitem [{\citenamefont {Tan}(1999)}]{Tan99}%
  \BibitemOpen
  \bibfield  {author} {\bibinfo {author} {\bibfnamefont {S.~M.}\ \bibnamefont
  {Tan}},\ }\href {http://stacks.iop.org/1464-4266/1/i=4/a=312} {\bibfield
  {journal} {\bibinfo  {journal} {J. Opt. B: Quantum Semiclass. Opt.}\ }\textbf
  {\bibinfo {volume} {1}},\ \bibinfo {pages} {424} (\bibinfo {year}
  {1999})}\BibitemShut {NoStop}%
\bibitem [{\citenamefont {Casabone}\ \emph
  {et~al.}(2013{\natexlab{a}})\citenamefont {Casabone}, \citenamefont {Stute},
  \citenamefont {Friebe}, \citenamefont {Brandst\"atter}, \citenamefont
  {Sch\"uppert}, \citenamefont {Blatt},\ and\ \citenamefont
  {Northup}}]{Casabone13}%
  \BibitemOpen
  \bibfield  {author} {\bibinfo {author} {\bibfnamefont {B.}~\bibnamefont
  {Casabone}}, \bibinfo {author} {\bibfnamefont {A.}~\bibnamefont {Stute}},
  \bibinfo {author} {\bibfnamefont {K.}~\bibnamefont {Friebe}}, \bibinfo
  {author} {\bibfnamefont {B.}~\bibnamefont {Brandst\"atter}}, \bibinfo
  {author} {\bibfnamefont {K.}~\bibnamefont {Sch\"uppert}}, \bibinfo {author}
  {\bibfnamefont {R.}~\bibnamefont {Blatt}}, \ and\ \bibinfo {author}
  {\bibfnamefont {T.~E.}\ \bibnamefont {Northup}},\ }\href {\doibase
  10.1103/PhysRevLett.111.100505} {\bibfield  {journal} {\bibinfo  {journal}
  {Phys. Rev. Lett.}\ }\textbf {\bibinfo {volume} {111}},\ \bibinfo {pages}
  {100505} (\bibinfo {year} {2013}{\natexlab{a}})}\BibitemShut {NoStop}%
\bibitem [{\citenamefont {Barros}\ \emph {et~al.}(2009)\citenamefont {Barros},
  \citenamefont {Stute}, \citenamefont {Northup}, \citenamefont {Russo},
  \citenamefont {Schmidt},\ and\ \citenamefont {Blatt}}]{Barros09}%
  \BibitemOpen
  \bibfield  {author} {\bibinfo {author} {\bibfnamefont {H.~G.}\ \bibnamefont
  {Barros}}, \bibinfo {author} {\bibfnamefont {A.}~\bibnamefont {Stute}},
  \bibinfo {author} {\bibfnamefont {T.~E.}\ \bibnamefont {Northup}}, \bibinfo
  {author} {\bibfnamefont {C.}~\bibnamefont {Russo}}, \bibinfo {author}
  {\bibfnamefont {P.~O.}\ \bibnamefont {Schmidt}}, \ and\ \bibinfo {author}
  {\bibfnamefont {R.}~\bibnamefont {Blatt}},\ }\href
  {http://stacks.iop.org/1367-2630/11/i=10/a=103004} {\bibfield  {journal}
  {\bibinfo  {journal} {New J. Phys.}\ }\textbf {\bibinfo {volume} {11}},\
  \bibinfo {pages} {103004} (\bibinfo {year} {2009})}\BibitemShut {NoStop}%
\bibitem [{\citenamefont {Stute}\ \emph {et~al.}(2012)\citenamefont {Stute},
  \citenamefont {Casabone}, \citenamefont {Brandst\"atter}, \citenamefont
  {Habicher}, \citenamefont {Schmidt}, \citenamefont {Northup},\ and\
  \citenamefont {Blatt}}]{Stute12a}%
  \BibitemOpen
  \bibfield  {author} {\bibinfo {author} {\bibfnamefont {A.}~\bibnamefont
  {Stute}}, \bibinfo {author} {\bibfnamefont {B.}~\bibnamefont {Casabone}},
  \bibinfo {author} {\bibfnamefont {B.}~\bibnamefont {Brandst\"atter}},
  \bibinfo {author} {\bibfnamefont {D.}~\bibnamefont {Habicher}}, \bibinfo
  {author} {\bibfnamefont {P.~O.}\ \bibnamefont {Schmidt}}, \bibinfo {author}
  {\bibfnamefont {T.~E.}\ \bibnamefont {Northup}}, \ and\ \bibinfo {author}
  {\bibfnamefont {R.}~\bibnamefont {Blatt}},\ }\href
  {http://arxiv.org/abs/1105.0579} {\bibfield  {journal} {\bibinfo  {journal}
  {Appl. Phys. B}\ }\textbf {\bibinfo {volume} {107}},\ \bibinfo {pages} {1145}
  (\bibinfo {year} {2012})}\BibitemShut {NoStop}%
\bibitem [{\citenamefont {S\o{}rensen}\ and\ \citenamefont
  {M\o{}lmer}(1999)}]{Sorensen99}%
  \BibitemOpen
  \bibfield  {author} {\bibinfo {author} {\bibfnamefont {A.}~\bibnamefont
  {S\o{}rensen}}\ and\ \bibinfo {author} {\bibfnamefont {K.}~\bibnamefont
  {M\o{}lmer}},\ }\href {\doibase 10.1103/PhysRevLett.82.1971} {\bibfield
  {journal} {\bibinfo  {journal} {Phys. Rev. Lett.}\ }\textbf {\bibinfo
  {volume} {82}},\ \bibinfo {pages} {1971} (\bibinfo {year}
  {1999})}\BibitemShut {NoStop}%
\bibitem [{\citenamefont {Sackett}\ \emph {et~al.}(2000)\citenamefont
  {Sackett}, \citenamefont {Kielpinski}, \citenamefont {King}, \citenamefont
  {Langer}, \citenamefont {Meyer}, \citenamefont {Myatt}, \citenamefont {Rowe},
  \citenamefont {Turchette}, \citenamefont {Itano}, \citenamefont {Wineland}
  \emph {et~al.}}]{Sackett00}%
  \BibitemOpen
  \bibfield  {author} {\bibinfo {author} {\bibfnamefont {C.}~\bibnamefont
  {Sackett}}, \bibinfo {author} {\bibfnamefont {D.}~\bibnamefont {Kielpinski}},
  \bibinfo {author} {\bibfnamefont {B.}~\bibnamefont {King}}, \bibinfo {author}
  {\bibfnamefont {C.}~\bibnamefont {Langer}}, \bibinfo {author} {\bibfnamefont
  {V.}~\bibnamefont {Meyer}}, \bibinfo {author} {\bibfnamefont
  {C.}~\bibnamefont {Myatt}}, \bibinfo {author} {\bibfnamefont
  {M.}~\bibnamefont {Rowe}}, \bibinfo {author} {\bibfnamefont {Q.}~\bibnamefont
  {Turchette}}, \bibinfo {author} {\bibfnamefont {W.}~\bibnamefont {Itano}},
  \bibinfo {author} {\bibfnamefont {D.}~\bibnamefont {Wineland}},  \emph
  {et~al.},\ }\href
  {http://iontrap.umd.edu/publications/archive/nature_404_256_sackett_4ions.pdf}
  {\bibfield  {journal} {\bibinfo  {journal} {Nature}\ }\textbf {\bibinfo
  {volume} {404}},\ \bibinfo {pages} {256} (\bibinfo {year}
  {2000})}\BibitemShut {NoStop}%
\bibitem [{\citenamefont {Schindler}\ \emph {et~al.}(2013)\citenamefont
  {Schindler}, \citenamefont {Nigg}, \citenamefont {Monz}, \citenamefont
  {Barreiro}, \citenamefont {Martinez}, \citenamefont {Wang}, \citenamefont
  {Quint}, \citenamefont {Brandl}, \citenamefont {Nebendahl}, \citenamefont
  {Roos} \emph {et~al.}}]{schindler13}%
  \BibitemOpen
  \bibfield  {author} {\bibinfo {author} {\bibfnamefont {P.}~\bibnamefont
  {Schindler}}, \bibinfo {author} {\bibfnamefont {D.}~\bibnamefont {Nigg}},
  \bibinfo {author} {\bibfnamefont {T.}~\bibnamefont {Monz}}, \bibinfo {author}
  {\bibfnamefont {J.~T.}\ \bibnamefont {Barreiro}}, \bibinfo {author}
  {\bibfnamefont {E.}~\bibnamefont {Martinez}}, \bibinfo {author}
  {\bibfnamefont {S.~X.}\ \bibnamefont {Wang}}, \bibinfo {author}
  {\bibfnamefont {S.}~\bibnamefont {Quint}}, \bibinfo {author} {\bibfnamefont
  {M.~F.}\ \bibnamefont {Brandl}}, \bibinfo {author} {\bibfnamefont
  {V.}~\bibnamefont {Nebendahl}}, \bibinfo {author} {\bibfnamefont {C.~F.}\
  \bibnamefont {Roos}},  \emph {et~al.},\ }\href@noop {} {\bibfield  {journal}
  {\bibinfo  {journal} {New J. Phys.}\ }\textbf {\bibinfo {volume} {15}},\
  \bibinfo {pages} {123012} (\bibinfo {year} {2013})}\BibitemShut {NoStop}%
\bibitem [{\citenamefont {Wineland}\ \emph {et~al.}(1998)\citenamefont
  {Wineland}, \citenamefont {Monroe}, \citenamefont {Itano}, \citenamefont
  {Leibfried}, \citenamefont {King},\ and\ \citenamefont
  {Meekhof}}]{Wineland98}%
  \BibitemOpen
  \bibfield  {author} {\bibinfo {author} {\bibfnamefont {D.~J.}\ \bibnamefont
  {Wineland}}, \bibinfo {author} {\bibfnamefont {C.}~\bibnamefont {Monroe}},
  \bibinfo {author} {\bibfnamefont {W.~M.}\ \bibnamefont {Itano}}, \bibinfo
  {author} {\bibfnamefont {D.}~\bibnamefont {Leibfried}}, \bibinfo {author}
  {\bibfnamefont {B.~E.}\ \bibnamefont {King}}, \ and\ \bibinfo {author}
  {\bibfnamefont {D.~M.}\ \bibnamefont {Meekhof}},\ }\href@noop {} {\bibfield
  {journal} {\bibinfo  {journal} {J. Res. Natl. Inst. Stand. Technol.}\
  }\textbf {\bibinfo {volume} {103}},\ \bibinfo {pages} {259} (\bibinfo {year}
  {1998})}\BibitemShut {NoStop}%
\bibitem [{\citenamefont {Stute}\ \emph {et~al.}(2013)\citenamefont {Stute},
  \citenamefont {Casabone}, \citenamefont {Brandst\"{a}tter}, \citenamefont
  {Friebe}, \citenamefont {Northup},\ and\ \citenamefont {Blatt}}]{Stute13}%
  \BibitemOpen
  \bibfield  {author} {\bibinfo {author} {\bibfnamefont {A.}~\bibnamefont
  {Stute}}, \bibinfo {author} {\bibfnamefont {B.}~\bibnamefont {Casabone}},
  \bibinfo {author} {\bibfnamefont {B.}~\bibnamefont {Brandst\"{a}tter}},
  \bibinfo {author} {\bibfnamefont {K.}~\bibnamefont {Friebe}}, \bibinfo
  {author} {\bibfnamefont {T.~E.}\ \bibnamefont {Northup}}, \ and\ \bibinfo
  {author} {\bibfnamefont {R.}~\bibnamefont {Blatt}},\ }\href {\doibase
  10.1038/nphoton.2012.358} {\bibfield  {journal} {\bibinfo  {journal} {Nature
  Photon.}\ }\textbf {\bibinfo {volume} {7}},\ \bibinfo {pages} {219} (\bibinfo
  {year} {2013})}\BibitemShut {NoStop}%
\bibitem [{\citenamefont {Chuang}\ and\ \citenamefont
  {Nielsen}(1997)}]{Chuang97}%
  \BibitemOpen
  \bibfield  {author} {\bibinfo {author} {\bibfnamefont {I.~L.}\ \bibnamefont
  {Chuang}}\ and\ \bibinfo {author} {\bibfnamefont {M.~A.}\ \bibnamefont
  {Nielsen}},\ }\href {\doibase 10.1080/09500349708231894} {\bibfield
  {journal} {\bibinfo  {journal} {J. Mod. Opt.}\ }\textbf {\bibinfo {volume}
  {44}},\ \bibinfo {pages} {2455} (\bibinfo {year} {1997})}\BibitemShut
  {NoStop}%
\bibitem [{\citenamefont {James}\ \emph {et~al.}(2001)\citenamefont {James},
  \citenamefont {Kwiat}, \citenamefont {Munro},\ and\ \citenamefont
  {White}}]{James01}%
  \BibitemOpen
  \bibfield  {author} {\bibinfo {author} {\bibfnamefont {D.~F.~V.}\
  \bibnamefont {James}}, \bibinfo {author} {\bibfnamefont {P.~G.}\ \bibnamefont
  {Kwiat}}, \bibinfo {author} {\bibfnamefont {W.~J.}\ \bibnamefont {Munro}}, \
  and\ \bibinfo {author} {\bibfnamefont {A.~G.}\ \bibnamefont {White}},\ }\href
  {\doibase 10.1103/PhysRevA.64.052312} {\bibfield  {journal} {\bibinfo
  {journal} {Phys. Rev. A}\ }\textbf {\bibinfo {volume} {64}},\ \bibinfo
  {pages} {052312} (\bibinfo {year} {2001})}\BibitemShut {NoStop}%
\bibitem [{\citenamefont {Cetina}\ \emph {et~al.}(2013)\citenamefont {Cetina},
  \citenamefont {Bylinskii}, \citenamefont {Karpa}, \citenamefont {Gangloff},
  \citenamefont {Beck}, \citenamefont {Ge}, \citenamefont {Scholz},
  \citenamefont {Grier}, \citenamefont {Chuang},\ and\ \citenamefont
  {Vuleti\'{c}}}]{Cetina13}%
  \BibitemOpen
  \bibfield  {author} {\bibinfo {author} {\bibfnamefont {M.}~\bibnamefont
  {Cetina}}, \bibinfo {author} {\bibfnamefont {A.}~\bibnamefont {Bylinskii}},
  \bibinfo {author} {\bibfnamefont {L.}~\bibnamefont {Karpa}}, \bibinfo
  {author} {\bibfnamefont {D.}~\bibnamefont {Gangloff}}, \bibinfo {author}
  {\bibfnamefont {K.~M.}\ \bibnamefont {Beck}}, \bibinfo {author}
  {\bibfnamefont {Y.}~\bibnamefont {Ge}}, \bibinfo {author} {\bibfnamefont
  {M.}~\bibnamefont {Scholz}}, \bibinfo {author} {\bibfnamefont {A.~T.}\
  \bibnamefont {Grier}}, \bibinfo {author} {\bibfnamefont {I.}~\bibnamefont
  {Chuang}}, \ and\ \bibinfo {author} {\bibfnamefont {V.}~\bibnamefont
  {Vuleti\'{c}}},\ }\href {http://stacks.iop.org/1367-2630/15/i=5/a=053001}
  {\bibfield  {journal} {\bibinfo  {journal} {New J. Phys.}\ }\textbf {\bibinfo
  {volume} {15}},\ \bibinfo {pages} {053001} (\bibinfo {year}
  {2013})}\BibitemShut {NoStop}%
\bibitem [{\citenamefont {H{\"a}ffner}\ \emph {et~al.}(2008)\citenamefont
  {H{\"a}ffner}, \citenamefont {Roos},\ and\ \citenamefont
  {Blatt}}]{Haeffner08}%
  \BibitemOpen
  \bibfield  {author} {\bibinfo {author} {\bibfnamefont {H.}~\bibnamefont
  {H{\"a}ffner}}, \bibinfo {author} {\bibfnamefont {C.}~\bibnamefont {Roos}}, \
  and\ \bibinfo {author} {\bibfnamefont {R.}~\bibnamefont {Blatt}},\ }\href
  {\doibase 10.1016/j.physrep.2008.09.003} {\bibfield  {journal} {\bibinfo
  {journal} {Phys. Rep.}\ }\textbf {\bibinfo {volume} {469}},\ \bibinfo {pages}
  {155} (\bibinfo {year} {2008})}\BibitemShut {NoStop}%
\bibitem [{\citenamefont {Reimann}\ \emph {et~al.}(2014)\citenamefont
  {Reimann}, \citenamefont {Alt}, \citenamefont {Kampschulte}, \citenamefont
  {Macha}, \citenamefont {Ratschbacher}, \citenamefont {Thau}, \citenamefont
  {Yoon},\ and\ \citenamefont {Meschede}}]{Reimann14}%
  \BibitemOpen
  \bibfield  {author} {\bibinfo {author} {\bibfnamefont {R.}~\bibnamefont
  {Reimann}}, \bibinfo {author} {\bibfnamefont {W.}~\bibnamefont {Alt}},
  \bibinfo {author} {\bibfnamefont {T.}~\bibnamefont {Kampschulte}}, \bibinfo
  {author} {\bibfnamefont {T.}~\bibnamefont {Macha}}, \bibinfo {author}
  {\bibfnamefont {L.}~\bibnamefont {Ratschbacher}}, \bibinfo {author}
  {\bibfnamefont {N.}~\bibnamefont {Thau}}, \bibinfo {author} {\bibfnamefont
  {S.}~\bibnamefont {Yoon}}, \ and\ \bibinfo {author} {\bibfnamefont
  {D.}~\bibnamefont {Meschede}},\ }\href@noop {} {\  (\bibinfo {year}
  {2014})},\ \Eprint {http://arxiv.org/abs/1408.5874 [quant-ph]}
  {arXiv:1408.5874 [quant-ph]} \BibitemShut {NoStop}%
\bibitem [{\citenamefont {Casabone}\ \emph
  {et~al.}(2013{\natexlab{b}})\citenamefont {Casabone}, \citenamefont {Stute},
  \citenamefont {Friebe}, \citenamefont {Brandst\"atter}, \citenamefont
  {Sch\"uppert}, \citenamefont {Blatt},\ and\ \citenamefont
  {Northup}}]{Casabone13Sup}%
  \BibitemOpen
  \bibfield  {author} {\bibinfo {author} {\bibfnamefont {B.}~\bibnamefont
  {Casabone}}, \bibinfo {author} {\bibfnamefont {A.}~\bibnamefont {Stute}},
  \bibinfo {author} {\bibfnamefont {K.}~\bibnamefont {Friebe}}, \bibinfo
  {author} {\bibfnamefont {B.}~\bibnamefont {Brandst\"atter}}, \bibinfo
  {author} {\bibfnamefont {K.}~\bibnamefont {Sch\"uppert}}, \bibinfo {author}
  {\bibfnamefont {R.}~\bibnamefont {Blatt}}, \ and\ \bibinfo {author}
  {\bibfnamefont {T.~E.}\ \bibnamefont {Northup}},\ }\href {\doibase
  10.1103/PhysRevLett.111.100505} {\bibfield  {journal} {\bibinfo  {journal}
  {Phys. Rev. Lett.}\ }\textbf {\bibinfo {volume} {111}},\ \bibinfo {pages}
  {100505} (\bibinfo {year} {2013}{\natexlab{b}})},\ \bibinfo {note}
  {{S}upplemental Material}\BibitemShut {NoStop}%
\end{thebibliography}%

\end{document}